\documentclass[final,english]{bullsrsl}

\usepackage[latin1]{inputenc}
\usepackage[T1]{fontenc}

\usepackage{natbib} 

\usepackage{graphicx}

\begin{document}
\title{Unveiling the Nature of C/2023 A3 (Tsuchinshan-ATLAS): A Multi-Technique Observational Approach}



  \author[affil={1,2}, corresponding]{Goldy}{Ahuja}
  \author[affil={1}]{Shashikiran}{Ganesh}
  \author[affil={3}]{Aravind}{Krishnakumar}
  \author[affil={4}]{Devendra K.}{Sahu}
  \author[affil={4}]{Thirupathi}{Sivarani}
  \author[affil={5}]{Vikrant K.}{Agnihotri}

\affiliation[1]{Physical Research Laboratory, Ahmedabad, 380009, India}
\affiliation[2]{Indian Institute of Technology Gandhinagar, Palaj, 382355, India}
\affiliation[3]{STAR Institute, University of Liege, Allee du 6 Aout 19c, 4000 Liege, Belgium}
\affiliation[4]{Indian Institute of Astrophysics, Koramangala, Bengaluru, Karnataka, 560034, India}
\affiliation[5]{Cepheid Observatory, Rawatbhata, Rajasthan, 323307, India}



\correspondance{goldy@prl.res.in}


\maketitle

\begin{abstract}
Comet C/2023 A3 (Tsuchinshan-ATLAS) is a non-periodic dynamically new Oort cloud comet that was discovered independently by Purple Mountain Observatory in China and Asteroid Terrestrial-impact Last Alert System (ATLAS) telescopes in South Africa. The comet passed perihelion at a distance of 0.39 AU on 27 September 2024. It was visible to the naked eye (the brightest since the comet C/1995 O1 (Hale-Bopp)) and was dubbed the great comet of 2024. In this work, we investigate the nature of this comet, which is moving in a hyperbolic orbit (e > 1), by analysing its composition using various observational techniques and tracing its orbital evolution through high-precision N-body simulations.
\end{abstract}





\section{Introduction}
Comets are basically the leftover materials after the formation of the solar system. They are either thrown out or, more generally, scattered into multiple reservoirs due to interactions with the Giant planets \citep{Tsiganis2005,Gomes2005,Morbidelli_2007,Walsh2011}. Initial classification was based on the orbital period of comets: bodies with an orbital period shorter than 200 years are called Short-Period Comets (hereafter, SPCs), and those with a longer orbital period are called Long-period comets (hereafter, LPCs). Further classifications are based on orbital parameters, such as using inclination, $(i)$, vs. semi-major axis, $(a)$, to find two subcategories for the SPCs, which are \textit{Jupiter Family Comets (JFCs)} and \textit{Halley Type comets (HTCs)} (see \citet{Levison_1996,Nesvorny_2017} for more details), while for the LPCs, the sub-classification was made to determine the populations of new and returning comets. \citet{Levison_1996} has given the limit of 10000 au for the semi-major axis of the object; if it is smaller than that, one believes the comet has been in the inner solar system before and is termed as \textit{returning comets}, while if it is larger than that, the comet is termed a \textit{new comet} or specifically termed as Dynamically New Comets (hereafter DNCs). However, recent simulations \citep[see][and references therein]{kroli_paper1,kroli_paper2} noted the need to revise the limit and added an additional condition: that the last perihelion, $q_{prev}$, should be greater than 15 au to be called DNCs.
Studying these DNCs will help us understand the chemical composition around the time when they first formed, as the bodies have not been in the inner solar system before, except when they were first formed, and this is practically the first time they are exposed to the sun. However, there is still a chance of galactic cosmic ray processing \citep{Gronoff_2012}. In this work, we examine the properties of the newly discovered comet C/2023 A3 (hereafter, comet A3) using different observational techniques and numerical orbital simulations. In section \ref{sec: Methodology}, we will briefly describe the methodology of the work, mainly observations (section \ref{sec: observations}), data reduction (section \ref{sec: data reduction}) and numerical modelling (section \ref{sec: numerical modelling}) and hence discuss the key results in section \ref{sec: Results}. In section \ref{sec: conclusion}, we will provide the summary and conclusion. 

\section{Observations, data reduction and Modelling details} \label{sec: Methodology}
We obtained low-resolution spectroscopic and broad-band photometric observations of long-period comet A3 using different instruments attached to different telescopes, including the $2~$m Himalayan Chandra Telescope (hereafter HCT), and the $0.132~$m Cepheid Observatory (hereafter CepO). The details are mentioned in the following subsections.

\subsection{Observations}\label{sec: observations}
Pre-perihelion spectroscopic observations were conducted from the $2~$m $f/9$ HCT (lat: 32.78 deg N; long 75.96 deg E; alt: 4500 m) using the Hanle Faint Object Spectrograph and Camera (HFOSC) instrument mounted at the Cassegrain focus \citep{HCT_paper}. The telescope, instrument, and observation details have been extensively covered in \citet{Ahuja_2025_MNRAS}. Since the emission of the daughter molecules in the comet's coma dominates in the optical region, i.e., 3800 \AA - 6900 \AA\ \citep{Ahuja_2025_MNRAS,Aravind_2025,Kumar_2016}, we used Grism 7, which provides a resolving power (R=$\lambda/\Delta \lambda $) of 1330 and wavelength coverage of 3800-6840 \AA.

The comet was observed with the long-slit (167l), which provides a slit length of 11 arcmin and a slit width of 1.92 arcsec, important for studying the varying column density and the radial distribution of the dust emission. The comet sky frame, used to remove the background, was acquired after moving the telescope 1 degree away from the photocenter in the direction of the comet's declination motion. Bias, continuum/flat frames, and FeAr lamp frames had been regularly acquired for bias correction, flat-fielding, and wavelength calibration. For flux calibration of the comets, we observed the standard stars using a wider slit (1340l), which provides a slit width of 15.4 arcsec and a length of 11 arcmin. It is capable of capturing the full flux of the star to avoid slit losses and, hence, any flux calibration issues in the comet spectra.

During post-perihelion, we conducted the spectroscopic observations from October 31$^{st}$ to November 04$^{th}$, 2024 using the 0.132~m-diameter, $f/4.6$ William Optics Fluorostar model refractor telescope and the photometric observations from November 06$^{th}-07^{th}$, 2024, using 0.073m $f/5.9$ William Optics Zenith star model refractor telescope at CepO (Lat: 24.55 deg N; Long: 75.34 deg E; Alt: 415 m). The spectroscopic observations were carried out using the ALPY600 instrument (R $\sim$ 600 at 6500 \AA) with SX-814 mono CCD. The CCD dimension considered is 1640 $\times$ 531 pixels, where each pixel size is 3.7 microns, which corresponds to the plate scale of 1.446 arcsec pixel$^{-1}$. The slit width is 23 microns, inclined at 22 degrees, giving an effective slit width of around 21.3 microns (or 8.322 arcsec), and the slit length is 3 mm (or 19.5 arcmin). We observed multiple standard stars listed in the ESO RA-ordered standards database, including HZ44 (sdO), BD+284211 (Op), Feige56 (B5p), HR7596 (A0III), and HR7950 (A1V), to flux-calibrate the comet spectrum. Since the spectrum is contaminated by the solar contribution, we observed different solar-analog stars, HD19445 (G2V) and HD186427 (G5V), to remove the comet's solar continuum.

\begin{table*}
\centering
\begin{minipage}{160mm}  
\centering
\caption{Observations details of Comet C/2023 (Tsuchinshan-ATLAS). Column "DTP" represents "Days To Perihelion"; "r" represents Heliocentric Distance; and "$\Delta$" is Geocentric distance}
\label{tab: observation_details}
\end{minipage}
\resizebox{\textwidth}{!}{%
\begin{tabular}{cccccccccc}
\hline
Date & DTP& r & $\Delta$ & Time & Telescope & Instrument/CCD & Observation & Phase & Airmass\\
 & (days) & (au) & (au) & (UT) & Facility & Used & Type  & Angle ($^\circ$) & \\
\hline
31-05-2024 & -119 & 2.33 & 1.79 & 16:50 & HCT & HFOSC & Spectroscopy (Gr7) & 24.17 & 1.56\\
15-06-2024 & -104 & 2.11 & 1.88 & 15:33 & HCT & HFOSC & Spectroscopy (Gr7) & 28.72 & 1.62\\
31-10-2024 & 34 & 0.92 & 0.96 & 13:14 & CepO & ALPY600 & Spectroscopy & 63.69 & 1.6\\
01-11-2024 & 35 & 0.94 & 0.99 & 13:20 & CepO & ALPY600 & Spectroscopy & 61.66 & 1.58\\
02-11-2024 & 36 & 0.96 & 1.03 & 13:13 & CepO & ALPY600 & Spectroscopy & 59.74 & 1.54\\
03-11-2024 & 37 & 0.98 & 1.07 & 13:14 & CepO & ALPY600 & Spectroscopy & 57.92 & 1.22\\
06-11-2024 & 40 & 1.03 & 1.17 & 14:11 & CepO & ATIK 383L+ & Photometry (B,V,R and I) & 52.97 & 2.35\\
07-11-2024 & 41 & 1.05 & 1.21 & 13:59 & CepO & ATIK 383L+ & Photometry (B,V,R and I) & 51.51 & 2.16\\
\hline
\end{tabular}
}
\end{table*}

The photometric observations were carried out using an Atik 383L+, which has a chip size of 3354 $\times$ 2529 pixels, where each pixel is of size 5.4 microns. To enhance sensitivity, the images are binned into 2 $\times$ 2. Thus, the effective plate scale is 5.17 arcsec pixel$^{-1}$. The observations were conducted on 06 to 07 November 2024 using the Bessel B, V, and R-band filters. The bandpass of the different filters is given in \citet{Bessell_UBV_2005}. The observation details for the comet A3 are mentioned in Table \ref{tab: observation_details}.

\subsection{Data Reduction}\label{sec: data reduction}
The spectroscopic data reduction was done using the self-scripted \textsc{Python} codes and different reduction packages from \textsc{IRAF} \citep[For more details, see section 3.1 of][]{Ahuja_2025_MNRAS}. Since the standard star in the HCT was observed with the wider slit, further corrections, such as slit correction and atmospheric dispersion, were not required, whereas in the spectroscopic observations from CepO, these corrections were needed. The details of these corrections are mentioned in the section 3.1 in \citet{Ahuja_2025_MNRAS}. 

Photometric data reduction was done using the Tycho tracker software \citep{2025epsc.conf.1222P}, which used Gaia DR3 for astrometry and the ATLAS catalogue for photometry \citep{ATLAS_1, Panstarrs_1, Gaiadr3_1}. The software itself automatically calibrates the comet coma using standard stars in the image field. It identifies the filter used in each frame and selects the corresponding catalog magnitudes for the field stars.

\subsection{Numerical Modelling}\label{sec: numerical modelling}
To understand its dynamical nature, we developed the in-house \textsc{Python} code using the REBOUND N-body package \citep{rebound}. The simulation setup is explained in detail in Section 2 in \citet{Ahuja_2025}. Using the method, we have created 100 statistical clones whose orbital parameters vary slightly based on the covariance matrix taken from NASA JPL \textsc{Horizons}. The hybrid integrator \texttt{Mercurius} \citep{reboundmercurius} was used to integrate the statistical clones of comet A3 backward in time for 10 million years. The time step used to integrate the system is $1/11$ of the innermost body, i.e., 8 days, which is optimal for studying changes in the different parameters and the comet's Cartesian velocities.
\begin{figure}
\centering
\includegraphics[width=0.48\textwidth]{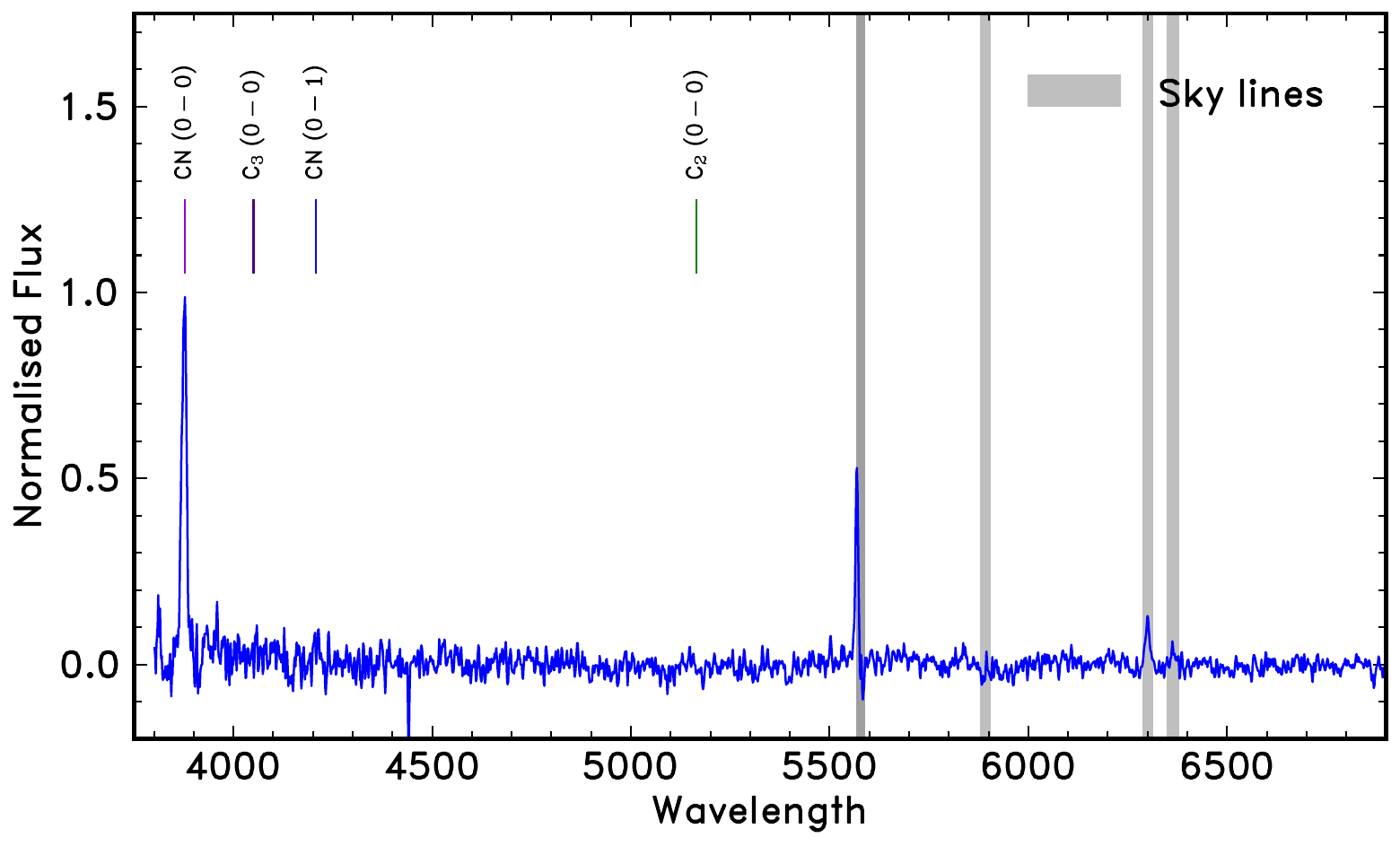}
\includegraphics[width=0.48\textwidth]{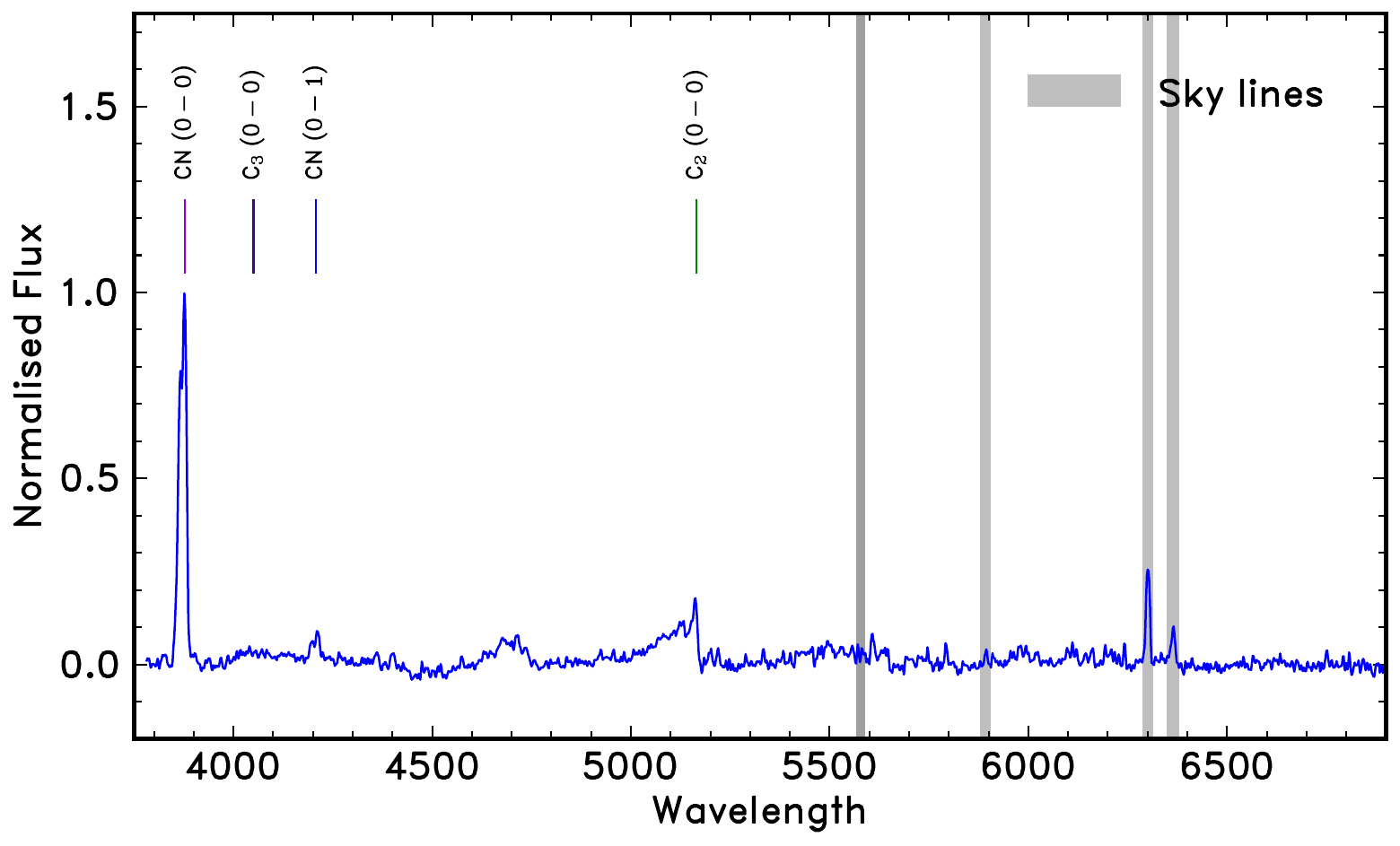}
\begin{minipage}{12cm}
\centering
\caption{Spectra of comet C/2023 A3: Pre-perihelion epoch using the HFOSC/HCT instrument(left panel);  Post-perihelion epoch using the ALPY600 spectrograph at CepO(right panel). The spectra show the stark difference in the observed molecular emissions before and after perihelion.}
\end{minipage}
\label{pre}
\end{figure}

\section{Results} \label{sec: Results}
Using the spectroscopic results, we analysed the pre-perihelion and post-perihelion spectra of the comet A3. The normalised spectra of the comet A3 are shown in Figure \ref{pre}.
As can be seen in the pre-perihelion spectrum (left panel of Figure \ref{pre}), the C$_2$, C$_3$ emissions were not present, and a clear detection of CN molecular emission was made, marking the first detection of CN in the comet A3 \citep{Ahuja_Goldy_ATel}. We calculated the production rates of different molecules using the single-aperture method and derived rate ratios such as C$_2$/CN and C$_3$/CN (see sections 4.2 of \citet{Ahuja_2025_MNRAS} and 2.3 of \citet{Aravind_2025} for more details). The production rate ratios are used to define the carbon class, such as whether the comet belongs to the typical or depleted class. The composition limit is defined by \citet{AHEARN1995223} based on observations of 85 comets. We have found that despite being an LPC, the comet is depleted in carbon composition, as also reported by \citet{Jehin_ATel_2024} and \citet{Cambianica_2025}. 

In the post-perihelion spectra (right panel of Figure \ref{pre}), we observed emissions from other daughter molecules apart from CN, yet we found a depleted carbon composition. The analysis confirmed that the comet is homogeneously depleted in carbon, which is very peculiar, as not many LPCs are depleted. The measured broadband colours of comet A3 are B$-$V $= 0.74 \pm 0.01$, V$-$R $= 0.48 \pm 0.01$, and R$-$I $= 0.45 \pm 0.01$. These colours are similar to the median colour of LPCs \citep{Jewitt_2015} and DNCs \citep{kulyk_2018}.

The numerical simulation results indicate that the last perihelion of comet A3 occurred around 9.5-10 MYr ago, when both gravitational and galactic tidal forces are taken into account. We also investigated the combined effects of gravitational and non-gravitational forces over a timescale of 1 Myr and found no significant difference relative to gravitational forces alone. When we put this comet under the Galactic Tidal force \citep[See section 4.1 of][and references therein]{kroli_paper1}, we found that the previous perihelion distance remained well outside the inner solar system, which confirms that the comet is a DNC.

\section{Summary and Conclusion}\label{sec: conclusion}
We have carried out photometric and spectroscopic observations as well as orbital evolution of the comet C/2023 A3, and the results are summarised below:
\begin{itemize}
    \item From spectroscopic observations at both perihelion and post-perihelion, we have found that comet A3 is depleted in carbon composition. The depleted nature is very peculiar, as it is found in only a handful of LPCs. This shows that the Oort Cloud is very heterogeneous. The post-perihelion broadband colour of the comet is found to be similar to the median colour of the LPCs.
    \item The N-body simulation results confirm that the comet had approached the Sun 9.5-10 Myr ago, and the previous perihelion of the comet is well outside of the solar system (q$_{prev} > 15$  au), which confirms that the comet A3 is a true DNC. 
    \item Further confirmations using the updated epoch and the other forces, such as stellar perturbations, may better constrain the results on the previous perihelion passage.
\end{itemize}




\begin{acknowledgments}
We acknowledge the local staff at the Mount Abu Observatory for their support. We thank the staff of IAO, Hanle and CREST, Hosakote, who made these observations possible. The facilities at IAO and CREST are operated by the Indian Institute of Astrophysics, Bangalore. Work at the Physical Research Laboratory is supported by the Department of Space, Govt. of India. The computations were performed on the Param Vikram-1000 High Performance Computing Cluster of the Physical Research Laboratory (PRL).

This work has made use of data from the Asteroid Terrestrial-impact Last Alert System (ATLAS) project. ATLAS is primarily funded to search for near-earth asteroids through NASA grants NN12AR55G, 80NSSC18K0284, and 80NSSC18K1575; byproducts of the NEO search include images and catalogues from the survey area.  The ATLAS science products have been made possible through the contributions of the University of Hawaii Institute for Astronomy, Queen's University Belfast, the Space Telescope Science Institute, and the South African Astronomical Observatory.
 
This work is a result of the bilateral Belgo-Indian projects on Precision Astronomical Spectroscopy for Stellar and solar system bodies, BIPASS, funded by the Belgian Federal Science Policy Office (BELSPO, Government of Belgium; BL$/$33$/$IN22\texttt{\_}BIPASS) and the International Division, Department of Science and Technology, (DST, Government of India; DST/INT/BELG/P-01/2021(G)).
\end{acknowledgments}

\begin{furtherinformation}

\begin{orcids}

  \orcid{0009-0008-1809-3256}{Goldy}{Ahuja}
  \orcid{0000-0002-7721-3827}{Shashikiran}{Ganesh}
  \orcid{0000-0002-8328-5667}{Aravind}{Krishnakumar}
  \orcid{0000-0002-6688-0800}{Devender}{K. Sahu}
  \orcid{0000-0003-0891-8994}{Thirupathi}{Sivarani}
\end{orcids}

\begin{authorcontributions}
G.A.: Conceptualisation, methodology, software, formal analysis, visualisation, writing-original draft.
S.G.: Supervision, writing- review and editing.
A.K.: formal analysis, writing- review and editing.
D.K.S.: Resources (observing proposal and observational support), writing- review and editing.
T.S.: Resources (observing proposal and observational support), writing- review and editing.
V.K.A.:  Resources (observational support), Investigation (spectroscopy and photometry), data analysis, writing- review and editing.
\end{authorcontributions}

\begin{conflictsofinterest}
The authors declare that there are no conflicts of interest regarding this work. 
\end{conflictsofinterest}

\end{furtherinformation}



%

\bibliographystyle{bullsrsl-en}

\bibliography{extra.bib}

\begin{thebibliography}{27}
\providecommand{\natexlab}[1]{#1}
\providecommand{\url}[1]{#1}
\providecommand{\urlprefix}{URL }

\bibitem[{A'Hearn et~al.(1995)A'Hearn, Millis, Schleicher, Osip and Birch}]{AHEARN1995223}
A'Hearn, M.~F., Millis, R.~C., Schleicher, D.~G., Osip, D.~J. and Birch, P.~V. (1995) The ensemble properties of comets: Results from narrowband photometry of 85 comets, 1976-1992.
\newblock Icarus, 118(2), 223--270.
\newblock \url{https://doi.org/https://doi.org/10.1006/icar.1995.1190}.

\bibitem[{Ahuja et~al.(2025)Ahuja, Aravind, Ganesh, Hmiddouch, Donckt, Jehin, Sahu and Sivarani}]{Ahuja_2025_MNRAS}
Ahuja, G., Aravind, K., Ganesh, S., Hmiddouch, S., Donckt, M.~V., Jehin, E., Sahu, D. and Sivarani, T. (2025) {Long-term monitoring of a dynamically new comet C/2020 V2 (ZTF)}.
\newblock Monthly Notices of the Royal Astronomical Society, 543(2), 1178--1195.
\newblock \url{https://doi.org/10.1093/mnras/staf1528}.

\bibitem[{{Ahuja} et~al.(2024){Ahuja}, {Aravind}, {Sahu}, {Jehin}, {Donckt}, {Hmiddouch}, {Ganesh} and {Sivarani}}]{Ahuja_Goldy_ATel}
{Ahuja}, G., {Aravind}, K., {Sahu}, D., {Jehin}, E., {Donckt}, M.~V., {Hmiddouch}, S., {Ganesh}, S. and {Sivarani}, T. (2024) {Molecular gas production rates of Comet C/2023 A3 (Tsuchinshan - ATLAS)}.
\newblock The Astronomer's Telegram, 16637, 1.

\bibitem[{Ahuja and Ganesh(2025)}]{Ahuja_2025}
Ahuja, G. and Ganesh, S. (2025) {Dynamical Simulation of the Interstellar Comet 3I/ATLAS}.
\newblock The Astrophysical Journal Letters, 995(1), L13.
\newblock \url{https://doi.org/10.3847/2041-8213/ae21cf}.

\bibitem[{{Aravind} et~al.(2025){Aravind}, {Jehin}, {Hmiddouch}, {Vander Donckt}, {Ganesh}, {Rousselot}, {Hardy}, {Sahu}, {Manfroid} and {Benkhaldoun}}]{Aravind_2025}
{Aravind}, K., {Jehin}, E., {Hmiddouch}, S., {Vander Donckt}, M., {Ganesh}, S., {Rousselot}, P., {Hardy}, P., {Sahu}, D., {Manfroid}, J. and {Benkhaldoun}, Z. (2025) {Ionic emission from and activity evolution in comet C/2020 F3 (NEOWISE): Insights from long-slit spectroscopy and photometry}.
\newblock Astronomy and Astrophysics, 701, A161.
\newblock \url{https://doi.org/10.1051/0004-6361/202554842}.

\bibitem[{{Bessell}(2005)}]{Bessell_UBV_2005}
{Bessell}, M.~S. (2005) {Standard Photometric Systems}.
\newblock Annual Review of Astronomy and Astrophysics, 43(1), 293--336.
\newblock \url{https://doi.org/10.1146/annurev.astro.41.082801.100251}.

\bibitem[{{Cambianica} et~al.(2025){Cambianica}, {Munaretto}, {Cremonese}, {Mura}, {La Forgia}, {Bizzocchi}, {Lazzarin}, {Puzzarini}, {Melosso}, {Lorenzi} and {Boschin}}]{Cambianica_2025}
{Cambianica}, P., {Munaretto}, G., {Cremonese}, G., {Mura}, A., {La Forgia}, F., {Bizzocchi}, L., {Lazzarin}, M., {Puzzarini}, C., {Melosso}, M., {Lorenzi}, V. and {Boschin}, W. (2025) {Pre-perihelion observations of the carbon-depleted comet C/2023 A3 (Tsuchinshan-ATLAS). Insights into CN production and molecular upper limits}.
\newblock Planetary and Space Science, 261, 106102.
\newblock \url{https://doi.org/10.1016/j.pss.2025.106102}.

\bibitem[{{Gaia Collaboration} et~al.(2023){Gaia Collaboration}, Vallenari, Brown, Prusti, de~Bruijne et~al.}]{Gaiadr3_1}
{Gaia Collaboration}, Vallenari, A., Brown, A.~G.~A., Prusti, T., de~Bruijne, J.~H.~J. et~al. (2023) Gaia data release 3. summary of the content and survey properties.
\newblock Astronomy and Astrophysics, 674, A1.
\newblock \url{https://doi.org/10.1051/0004-6361/202243940}.

\bibitem[{Gomes et~al.(2005)Gomes, Levison, Tsiganis and Morbidelli}]{Gomes2005}
Gomes, R., Levison, H.~F., Tsiganis, K. and Morbidelli, A. (2005) Origin of the cataclysmic late heavy bombardment period of the terrestrial planets.
\newblock Nature, 435(7041), 466--469.
\newblock \url{https://doi.org/10.1038/nature03676}.

\bibitem[{{Gronoff} et~al.(2020){Gronoff}, {Maggiolo}, {Cessateur}, {Moore}, {Airapetian}, {De Keyser}, {Dhooghe}, {Gibbons}, {Gunell}, {Mertens}, {Rubin} and {Hosseini}}]{Gronoff_2012}
{Gronoff}, G., {Maggiolo}, R., {Cessateur}, G., {Moore}, W.~B., {Airapetian}, V., {De Keyser}, J., {Dhooghe}, F., {Gibbons}, A., {Gunell}, H., {Mertens}, C.~J., {Rubin}, M. and {Hosseini}, S. (2020) {The Effect of Cosmic Rays on Cometary Nuclei. I. Dose Deposition}.
\newblock The Astrophysical Journal, 890(1), 89.
\newblock \url{https://doi.org/10.3847/1538-4357/ab67b9}.

\bibitem[{{Jehin} et~al.(2024){Jehin}, {Donckt}, {Hmiddouch} and {Manfroid}}]{Jehin_ATel_2024}
{Jehin}, E., {Donckt}, M.~V., {Hmiddouch}, S. and {Manfroid}, J. (2024) {TRAPPIST bright comets production rates: 13P/Olbers, C/2023 A3 (Tsuchinshan - ATLAS) and C/2021 S3 (PanSTARRS)}.
\newblock The Astronomer's Telegram, 16705, 1.

\bibitem[{Jewitt(2015)}]{Jewitt_2015}
Jewitt, D. (2015) Color systematics of comets and related bodies*.
\newblock The Astronomical Journal, 150(6), 201.
\newblock \url{https://doi.org/10.1088/0004-6256/150/6/201}.

\bibitem[{{Kostov} and {Bonev}(2018)}]{Panstarrs_1}
{Kostov}, A. and {Bonev}, T. (2018) {Transformation of Pan-STARRS1 gri to Stetson BVRI magnitudes. Photometry of small bodies observations.}
\newblock Bulgarian Astronomical Journal, 28, 3.
\newblock \url{https://doi.org/10.48550/arXiv.1706.06147}.

\bibitem[{{Kr{\'o}likowska}(2014)}]{kroli_paper2}
{Kr{\'o}likowska}, M. (2014) {Warsaw Catalogue of cometary orbits: 119 near-parabolic comets}.
\newblock Astronomy and Astrophysics, 567, A126.
\newblock \url{https://doi.org/10.1051/0004-6361/201323263}.

\bibitem[{Kr{\'o}likowska and Dybczy{\'n}ski(2010)}]{kroli_paper1}
Kr{\'o}likowska, M. and Dybczy{\'n}ski, P.~A. (2010) {Where do long-period comets come from? 26 comets from the non-gravitational Oort spike}.
\newblock Monthly Notices of the Royal Astronomical Society, 404(4), 1886--1902.
\newblock \url{https://doi.org/10.1111/j.1365-2966.2010.16403.x}.

\bibitem[{{Kulyk} et~al.(2018){Kulyk}, {Rousselot}, {Korsun}, {Afanasiev}, {Sergeev} and {Velichko}}]{kulyk_2018}
{Kulyk}, I., {Rousselot}, P., {Korsun}, P.~P., {Afanasiev}, V.~L., {Sergeev}, A.~V. and {Velichko}, S.~F. (2018) {Physical activity of the selected nearly isotropic comets with perihelia at large heliocentric distance}.
\newblock Astronomy and Astrophysics, 611, A32.
\newblock \url{https://doi.org/10.1051/0004-6361/201731529}.

\bibitem[{Levison(1996)}]{Levison_1996}
Levison, H.~F. (1996) {Comet Taxonomy}.
\newblock Completing the Inventory of the Solar System, Astronomical Society of the Pacific Conference Proceedings, 107, 173--191.

\bibitem[{Morbidelli et~al.(2007)Morbidelli, Tsiganis, Crida, Levison and Gomes}]{Morbidelli_2007}
Morbidelli, A., Tsiganis, K., Crida, A., Levison, H.~F. and Gomes, R. (2007) Dynamics of the giant planets of the solar system in the gaseous protoplanetary disk and their relationship to the current orbital architecture.
\newblock The Astronomical Journal, 134(5), 1790.
\newblock \url{https://doi.org/10.1086/521705}.

\bibitem[{Nesvorn{\'y} et~al.(2017)Nesvorn{\'y}, Vokrouhlick{\'y}, Dones, Levison, Kaib and Morbidelli}]{Nesvorny_2017}
Nesvorn{\'y}, D., Vokrouhlick{\'y}, D., Dones, L., Levison, H.~F., Kaib, N. and Morbidelli, A. (2017) Origin and evolution of short-period comets.
\newblock The Astrophysical Journal, 845(1), 27.
\newblock \url{https://doi.org/10.3847/1538-4357/aa7cf6}.

\bibitem[{{Parrott} and {Zakraj{\v{s}}ek}(2025)}]{2025epsc.conf.1222P}
{Parrott}, D. and {Zakraj{\v{s}}ek}, J. (2025) {Tycho Tracker: A Comprehensive All-in-One Tool for Comet Photometry}.
\newblock In EPSC-DPS Joint Meeting 2025, vol. 2025, pp. EPSC--DPS2025--1222.
\newblock \url{https://doi.org/10.5194/epsc-dps2025-1222}.

\bibitem[{{Prabhu}(2014)}]{HCT_paper}
{Prabhu}, T.~P. (2014) {Indian Astronomical Observatory, Leh-Hanle}.
\newblock Proceedings of the Indian National Science Academy Part A, 80(4), 887--912.
\newblock \url{https://doi.org/10.16943/ptinsa/2014/v80i4/55174}.

\bibitem[{{Rein} et~al.(2019){Rein}, {Hernandez}, {Tamayo}, {Brown}, {Eckels}, {Holmes}, {Lau}, {Leblanc} and {Silburt}}]{reboundmercurius}
{Rein}, H., {Hernandez}, D.~M., {Tamayo}, D., {Brown}, G., {Eckels}, E., {Holmes}, E., {Lau}, M., {Leblanc}, R. and {Silburt}, A. (2019) {Hybrid symplectic integrators for planetary dynamics}.
\newblock Monthly Notices of the Royal Astronomical Society, 485(4), 5490--5497.
\newblock \url{https://doi.org/10.1093/mnras/stz769}.

\bibitem[{{Rein} and {Liu}(2012)}]{rebound}
{Rein}, H. and {Liu}, S.~F. (2012) {REBOUND: an open-source multi-purpose N-body code for collisional dynamics}.
\newblock Astronomy and Astrophysics, 537, A128.
\newblock \url{https://doi.org/10.1051/0004-6361/201118085}.

\bibitem[{{Tonry} et~al.(2018){Tonry}, {Denneau}, {Flewelling}, {Heinze}, {Onken}, {Smartt}, {Stalder}, {Weiland} and {Wolf}}]{ATLAS_1}
{Tonry}, J.~L., {Denneau}, L., {Flewelling}, H., {Heinze}, A.~N., {Onken}, C.~A., {Smartt}, S.~J., {Stalder}, B., {Weiland}, H.~J. and {Wolf}, C. (2018) {The ATLAS All-Sky Stellar Reference Catalog}.
\newblock The Astrophysical Journal, 867(2), 105.
\newblock \url{https://doi.org/10.3847/1538-4357/aae386}.

\bibitem[{Tsiganis et~al.(2005)Tsiganis, Gomes, Morbidelli and Levison}]{Tsiganis2005}
Tsiganis, K., Gomes, R., Morbidelli, A. and Levison, H.~F. (2005) Origin of the orbital architecture of the giant planets of the solar system.
\newblock Nature, 435(7041), 459--461.
\newblock \url{https://doi.org/10.1038/nature03539}.

\bibitem[{Venkataramani et~al.(2016)Venkataramani, Ghetiya, Ganesh, Joshi, Agnihotri and Baliyan}]{Kumar_2016}
Venkataramani, K., Ghetiya, S., Ganesh, S., Joshi, U.~C., Agnihotri, V.~K. and Baliyan, K.~S. (2016) Optical spectroscopy of comet c/2014 q2 (lovejoy) from the mount abu infrared observatory.
\newblock Monthly Notices of the Royal Astronomical Society, 463(2), 2137--2144.
\newblock \url{https://doi.org/10.1093/mnras/stw1820}.

\bibitem[{Walsh et~al.(2011)Walsh, Morbidelli, Raymond, O'Brien and Mandell}]{Walsh2011}
Walsh, K.~J., Morbidelli, A., Raymond, S.~N., O'Brien, D.~P. and Mandell, A.~M. (2011) A low mass for mars from jupiter's early gas-driven migration.
\newblock Nature, 475(7355), 206--209.
\newblock \url{https://doi.org/10.1038/nature10201}.

\end{thebibliography}

\end{document}